\documentclass[doublecol]{epl2} 
\usepackage{amssymb}
\usepackage{graphicx}
\usepackage{dcolumn}
\usepackage{amsmath}
\usepackage{bm}
 
\usepackage{epsfig}

\newtheorem{theorem}{Theorem}

\newenvironment{proof}[1][Proof]{\begin{trivlist}
\item[\hskip \labelsep {\bfseries #1}]}{\end{trivlist}}
\newenvironment{definition}[1][Definition]{\begin{trivlist}
\item[\hskip \labelsep {\bfseries #1}]}{\end{trivlist}}

\newenvironment{remark}[1][Remark]{\begin{trivlist}
\item[\hskip \labelsep {\bfseries #1}]}{\end{trivlist}}

\newcommand{\qed}{\nobreak \ifvmode \relax \else
      \ifdim\lastskip<1.5em \hskip-\lastskip
      \hskip1.5em plus0em minus0.5em \fi \nobreak
      \vrule height0.75em width0.5em depth0.25em\fi}

\def\defi{{\buildrel \;def\; \over =}}
\newcommand{\be}{\begin{equation}}
\newcommand{\ee}{\end{equation}}

\newcommand{\media}[1]{\langle #1 \rangle}

\title{Mean-field models with short-range correlations}
\shorttitle{Mean field models with short-range correlations}
\author{M. Ostilli \inst{1,2}}
\institute{                    
  \inst{1} Dept. of Computational \& Theoretical Sciences,
IIUM,
Kuantan, Pahang, Malaysia \\
  \inst{2} Statistical Mechanics and Complexity Center (SMC), INFM-CNR SMC, Rome, Italy
} 


\abstract{Given an arbitrary finite dimensional Hamiltonian $H_0$,
we consider the model $H=H_0+\Delta H$,
where $\Delta H$ is a generic fully connected interaction.
By using the strong law of large numbers we easily prove that, 
for all such models, a generalized Curie-Weiss mean-field equation holds.
Unlike traditional mean-field models the term $H_0$ gives rise to
short-range correlations and, furthermore, when $H_0$ has negative couplings,
first-order phase transitions and inverse transition phenomena may take place
even when only two-body interactions are present. The dependence from a non uniform
external field and finite size effects are also explicitly calculated. 
Partially, these results were derived long ago by using min-max principles
but remained almost unknown.}

\begin{document}

\maketitle

\email{massimo.ostilli@roma1.infn.it}

\section{Introduction}
\label{intro} 
Mean field theories are usually introduced as methods able
to solve exactly models for which correlations are negligible: 
\begin{eqnarray}
\label{decorr}
\media{\sigma_i\sigma_j}\simeq\media{\sigma_i}\media{\sigma_j}, \quad i\neq j,
\end{eqnarray}
where $\media{\sigma_i}$ is the thermal average of a generic variable $\sigma_i$,
and the equality holds in the thermodynamic limit within a pure state \cite{Sinai}.
The basic prototype of these models is the classical mean-field Ising model
in which every spin $\sigma_i=\pm 1$, is interacting with any other spin, the Hamiltonian being given 
by the sum of the interactions $\sigma_i\sigma_j$ over the fully connected (FC) graph 
in the presence of a field $h$ \cite{Temperley}:
\begin{eqnarray}
\label{HCW}
H\defi -\frac{J}{N}\sum_{i<j} \sigma_{i}\sigma_{j}+h\sum_i\sigma_i.
\end{eqnarray}
This model is exactly solvable and, in the thermodynamic limit $N\to\infty$, 
the Gibbs-Boltzmann average $\media{\mathop{O}}$ of any quantity $\mathop{O}$, 
can be calculated starting from the knowledge of the order parameter $m=\sum_i\media{\sigma_i}/N$, 
which satisfies the famous Curie-Weiss mean-field equation ($\beta=1/T$):
\begin{eqnarray}
\label{mCW}
m=\tanh(\beta J m +\beta h).
\end{eqnarray}
As is well known, from (\ref{mCW}) it follows that the system (\ref{HCW}) has a second-order phase transition
at the critical point $\beta_c J=1$ near which $m$ develops a critical behavior with
the classical critical exponents $\beta=1/2$, $\gamma=1$ and $\delta=1/3$. 

A first method (M1) to prove Eq. (\ref{mCW}),
consists in using (\ref{decorr}) in (\ref{HCW}) for $\sigma_i\sigma_j\simeq \sigma_i m+\sigma_j m-m^2$ and then, 
\textit{a posteriori}, to prove that (\ref{decorr})
is indeed true. 
A second method (M2) to prove Eq. (\ref{mCW}), 
is based on the observation that, up to finite size effects, $H$ is a function 
of the sole total magnetization  
\begin{eqnarray}
\label{HM}
H=H(M), \quad M=\sum_i\sigma _i.
\end{eqnarray}
Eq. (\ref{HM}) implies that the sum over the 
spin-configurations of $e^{-\beta H(M)}$ can be calculated as a sum over the spin-configurations
with $M$ fixed
of $e^{-\beta H(M)}\mathcal{N}(M)$, where 
$\mathcal{N}(M)$ 
is the number of spin-configurations with given $M$.  

In this Letter we show that, besides M1 and M2, 
there exist two other methods based on the Gaussian Transformation (GT) and on the
Strong Law of Large Numbers (SLLN), respectively, that turn out to be much more
powerful than M1 and M2 since, neither for the GT, nor
for the SLLN, the conditions (\ref{decorr}) or (\ref{HM}) are necessary.
{By using GT's or SLLN's we obtain the following result:
given an arbitrary Hamiltonian $H_0=H_0(h)$ in the presence of a uniform external field $h$,
(later on we shall consider also non uniform fields),
the order parameter $m$ of the model 
\begin{eqnarray}
\label{HIntro}
H=H_0(h)+\Delta H 
\end{eqnarray}
with $\Delta H$ a generic  
FC interaction (see Eqs. (\ref{Cond1}) or (\ref{VecCond1}) for the general definition),
satisfies a generalized version of the Curie-Weiss mean-field Eq. (\ref{mCW}), of the kind:
\begin{eqnarray}
\label{m}
m=m_0(\mathrm{parameters};\beta U(m)+\beta h),
\end{eqnarray}
where: $m_0(\mathrm{parameters}; \beta h)$ is the order-parameter of the system
$H_0(h)$ at $T=1/\beta$; 
``parameters'' , in the first argument of $m_0$, 
refers to the other parameters contained in $H_0$ (the couplings and the features of the graph on which $H_0$ is built on) 
rescaled by $\beta$; and $U$ is a function (a potential) determined by $\Delta H$.
Furthermore, for large system size $N$, all the correlation functions
$C$ of the system governed by $H$, are simply related to the correlation functions $C_0$ 
of the system governed by $H_0$ by relations of the kind:
\begin{eqnarray}
\label{m1}
C=C_0(\mathrm{parameters};\beta U(m)+\beta h) +\mathrm{Finite~Size~Effects}.
\end{eqnarray} }
We will see that, when $\Delta H \propto M^p$, with $p$ integer, 
Eqs. (\ref{m}) and (\ref{m1}) can be proved by using a finite number of GT's without any further condition,
whereas, for the most general FC $\Delta H$, we will need to use SLLN valid under
mild bounds on the connected correlation functions (CCF).

Of course, the exact solution for $H_0$
remains in general a formidable task,
however, by using Eqs. (\ref{m}-\ref{m1}) we see that, 
even if we do not have the function $m_0(\mathrm{parameters};\beta h)$,
but we know at least some of its properties,
we can still use these properties to derive important exact relations
of the model governed by $H$. Eqs. (\ref{m}-\ref{m1}) imply
that the addition of a FC interaction is a perturbation that sends \textit{almost all}
systems, finite dimensional or not, toward a strict class of mean-field models where, however, unlike ordinary
mean-field models for which Eqs. (\ref{decorr}) or (\ref{HM}) are true, 
correlations survive even in the thermodynamic limit. 
Moreover, when $H_0$ has negative couplings,
first-order phase transitions and inverse transition phenomena 
may take place, even when only two-body interactions are present.

{The general result expressed by Eq. (\ref{m}) was first
derived long ago within a min-max principle by Bogoliubov Jr. for the
case in which $\Delta H$ is quadratic \cite{Bogoliubov}, 
and then extended to a much wider class by \cite{Perk}, including the quantum case.   
A rich scenario of phase transitions was also found in \cite{Kaufman} for
a simple competitive Ising model on the Bethe lattice, where a GT was used. 
But, quite surprisingly, 
these results turn out to be almost totally unknown and neglected.
On the other hand, in recent years, a renewed attention to models having both short- and long-range
interactions has been seen 
within the context of random small-world networks \cite{Watts}, in \cite{Hastings2,Skantzos,SW}. 
In this class of models, the FC interaction can be seen as the limit case in which the average connectivity, $c$, 
of the small-world network approaches the system size,
$c\to N$, with the coupling rescaled as $J\to J/N$. In such a limit, the equations we introduced in
\cite{SW} for the Ising model 
become exact in all the phase diagram and constitute a particular case (the Ising case),
of the general result expressed by Eqs. (\ref{m}-\ref{m1}). 
This implies, in particular,
that the technique used in \cite{SW} can be readily extended to analyze much more general random models.

The main scope of this Letter is to prove Eqs. (\ref{m}-\ref{m1}) in the most general case,
including the presence of a non uniform external field,
by using a very compact proof based on SLLN, and to provide the explicit expression 
for the $\mathop{O}(1/N)$ Finite Size Effects in Eq. (\ref{m1}).
Apart from the Ising case (already analyzed in detail in \cite{SW}), 
a complete analysis of what Eqs. (\ref{m}-\ref{m1}) imply 
for each class of model is far beyond the aim of the present work.
However, since the mean-field limit 
constitutes one of the most important approximations to approach realistic models, including disordered models, 
we believe that Eqs. (\ref{m}-\ref{m1}) will stimulate future investigations of specific models.
Moreover, in \cite{Rikvold} it has been recently shown that certain 
short-range models having additional kinetic degrees of freedom,
behave effectively as special mean-field models obeying Eqs. (\ref{m}-\ref{m1}).}

For the sake of clarity and rigor, the general result derived by using the SLLN is given
in the form of two theorems devoted to scalar and vector models, respectively.

\section{Models via Gaussian Transformations}
In this Section we will show how to derive Eqs. (\ref{m}) and (\ref{m1}) by using the
GT $\int dx \exp(-ax^2/2+bx)\propto \exp(b^2/(2a))$, or its generalizations to more components.
Here, we will avoid the full generality, and we will follow the strategy to illustrate concrete models step by step.

\subsection{Ising case}
Let us consider first a regular Ising model with a single coupling $J_0$ and with no external field
\begin{eqnarray}
\label{H0}
H_0=-J_0\sum_{(i,j)_0}\sigma_i\sigma_j 
\end{eqnarray}
where $(i,j)_0$ stands for first neighbors with respect to the lattice on which $H_0$ is built on.
Let $f_0(\beta J_0;\{\beta h_i\})$ and $m_0(\beta J_0;\{\beta h_i\})$ be
the free energy density and the average magnetization for the system governed by $H_0$
immersed in an arbitrary external field $\{h_i\}$. More in general, in the following, a symbol like 
$A_0(\{h_i\})$, will refer to the system governed by $H_0$ in the presence of the external field $\{h_i\}$.
Let us consider now the following Hamiltonian in the presence of a uniform external field $h$ \cite{Oitmaa}:
\begin{eqnarray}
\label{H}
H=-J_0\sum_{(i,j)_0}\sigma_i\sigma_j -\frac{J}{N}\sum_{i<j}\sigma_i\sigma_j- h\sum_i \sigma_i
\end{eqnarray}
In order to calculate the partition function $Z$ of this system, we need first to use a GT after which we get
\begin{eqnarray}
\label{Z}
&& Z=\sum_{\{\sigma_i\}}e^{-\beta H}=\frac{\sqrt{\beta J}}{\sqrt{2\pi N}}\int du ~e^{-\frac{\beta J N u^2}{2}}
\nonumber \\ && \times 
\sum_{\{\sigma_i\}} e^{\beta J_0\sum_{(i,j)_0}\sigma_i\sigma_j +\left(\beta J+\beta h\right)\sum_{i}\sigma_i+\mathop{O}(1)},
\end{eqnarray}  
where we have used the fact that, for finite $N$, we can exchange the integral with the sum~
\footnote{
Here we have supposed that $J>0$.
The procedure for $J<0$ differs from the other just for a rotation of $\pi/2$ in the
complex $u$ plane and leads to the same result one can obtain
by analytically continue the equations derived for $J>0$ to the region $J<0$.}. 
From Eq. (\ref{H0}), we see that the partition function involved in Eq. (\ref{Z})
is the partition function for $H_0$ with $\beta h_i\equiv \beta h+\beta J u$, therefore, 
by using $-\beta f_0(\beta J_0;\beta h)=\log(Z_0(\{h_i\}))|_{\{h_i=h\}}$, we obtain
\begin{eqnarray}
\label{Z1}
Z\propto\int du e^{-NL(u)}, \quad
L(u)=\frac{\beta J u^2}{2}+\beta f_0(\beta J_0;\beta J u).
\end{eqnarray}  
From Eqs. (\ref{Z1}), by using the saddle point method 
we find that, for $N\to\infty$, the free energy density $f$ for $H$ is 
\begin{eqnarray}
\label{ff}
&& \beta f=\frac{\beta J m^2}{2}+\beta f_0\left(\beta J_0;\beta Jm+\beta h\right), \quad \mathrm{where} ~~~~~~~~ \\
\label{mm}
&& m=m_0\left(\beta J_0;\beta J m +\beta h\right),
\end{eqnarray}
and, by derivating $-\beta f$ with respect to $\beta h$, it is immediate to verify that
$m$, solution of Eq. (\ref{mm}), coincides with the order parameter for $H$: $m=\media{\sigma}$.
More in general, if we use $\beta f_0(\beta J_0;\{\beta h_i\})$ and $m_{0;i}(\beta J_0;\{\beta h_i\})$,
as functions of an arbitrary external field $\{h_i\}$,
by following almost identical steps as above, with
\begin{eqnarray}
\label{Hg}
&& H=-J_0\sum_{(i,j)_0}\sigma_i\sigma_j -\frac{J}{N}\sum_{i<j}\sigma_i\sigma_j- \sum_i h_i\sigma_i, ~
\mathrm{we~get} \nonumber ~~~~~~\\
\label{ffg}
&& \beta f=\frac{\beta J m^2}{2}+\beta f_0\left(\beta J_0;\{\beta Jm+\beta h_i\}\right), ~
\mathrm{where} \nonumber ~~~~~~\\
\label{mmg}
&& m=\lim_{N\to\infty}\frac{1}{N}\sum_i m_{0i}\left(\beta J_0;\{\beta J m +\beta h_i\}\right).
\end{eqnarray}
Eqs. (\ref{mm}) or (\ref{mmg}) show that the system governed by $H$ obeys, at least for positive couplings,
a generalized Curie-Weiss mean-field equation with classical critical indices \cite{SW}.
However, by derivating $f$ with respect to $h_i$ and $h_j$ we get
\begin{eqnarray}
\label{CCg}
C_{i,j}^{(2)}=C_{0;i,j}^{(2)}\left(\beta J_0;\{\beta J m +\beta h_i\}\right),
\end{eqnarray}
where $C_{i,j}^{(2)}$ and $C_{0;i,j}^{(2)}$ are the correlation functions for $H$ and $H_0$, respectively,
in the thermodynamic limit. Similarly, it is easy to see that the transformation rule as indicated by 
Eq. (\ref{CCg}), holds for correlation functions of any order, connected or not. We see therefore that
if the system governed by $H_0$ has non zero CCF, CCF for the system governed by $H$ 
will be non zero too; the latter is a feature which is absent in ordinary
mean-field theories.

\subsection{$p$-spins case}
Let us now see how the previous result generalizes to the case of $p$-spins Hamiltonians with $p>2$.
Let us consider the simplest cases with $p=3$ and $p=4$ but with $H_0$ a $p'$-spins system with $p'$ 
arbitrary:
\begin{eqnarray}
\label{Hgp}
&& H=-J_0\sum_{(i_1,\ldots,i_{p'})_0}\sigma_{i_1}\cdots\sigma_{i_{p'}} \nonumber \\ &&
-\frac{(p-1)!J}{N^{p-1}}\sum_{i_1<\ldots<i_p} \sigma_{i_1}\cdots \sigma_{i_p} - h \sum_i \sigma_i,
\end{eqnarray}
It is worth to remind that, for $p'>2$, $J_0$ also must scale with $N$, but here we do not need to specify it.
If $p=4$, after two GT's we arrive easily to the following expression:
\begin{eqnarray}
\label{Z1p4}
&& Z\propto \int du dv ~e^{-NL(u,v)}, \quad \mathrm{where~for~}p=4 \nonumber ~~~\\
&& L(u,v)=\frac{\beta J u^2}{4}+\frac{\beta J v^2}{2u}+\beta f_0\left(\beta J_0;\beta J v+\beta h\right).
\nonumber
\end{eqnarray}  
By saddle point method and by introducing the new variable $m=v^{1/3}$, we get the saddle point equation
\begin{eqnarray}
\label{mp4}
m=m_0^{(p')}(\beta J_0;\beta J m^{3}+\beta h), \quad p=4
\end{eqnarray}  
where $m_0^{(p')}(\beta J_0;\beta h)$ is the order parameter for $H_0$ as a function of an homogeneous
external field $h$: $m_0^{(p')}(\beta J_0;\beta h)=\media{\sigma}_0$. 
Finally, by derivating with respect to $\beta h$ it is easy to check that, $m$, solution
of Eq. (\ref{mp4}), is the order parameter for $H$: $m=\media{\sigma}$.
Similarly, for $p=3$, by using two GT's we arrive to a skew expression for $Z$
\begin{eqnarray}
\label{Z1p3}
&& Z\propto \int du dv ~e^{-NL(u,v)}, \quad \mathrm{where~for~} p=3 \nonumber ~~~\\
&& L(u,v)=\frac{\beta J v^2}{12u}+\beta f_0\left(\beta J_0;\frac{\beta J v}{3}-\frac{\beta J u^2}{12}+\beta h\right).
\nonumber
\end{eqnarray}  
By saddle point method and by introducing the new variable $m=u/2$, we get the saddle point equation
\begin{eqnarray}
\label{mp5}
m=m_0^{(p')}(\beta J_0;\beta J m^{2}+\beta h), \quad p=3,
\end{eqnarray}  
where $m$ is the order parameter for $H$: $m=\media{\sigma}$.
In general, in agreement with Eq. (\ref{m}), for any $p$ and $p'$:
\begin{eqnarray}
\label{mpg}
m=m_0^{(p')}(\beta J_0;\beta J m^{p-1}+\beta h). 
\end{eqnarray}  

\subsection{Potts Model case}
Let us now consider the Potts model, \textit{i.e.}, the model where each variable $\sigma$
can take $q$ values: $\sigma=0,\ldots,q-1$ with Hamiltonian 
\begin{eqnarray}
\label{HgPotts}
H=H_0(\{\sigma_i\})-\frac{J}{N}\sum_{i<j}\delta(\sigma_i,\sigma_j)-\sum_{\sigma} h^{(\sigma)} N^{(\sigma)},
\end{eqnarray}
where $\delta(\sigma,\sigma')$ is the Kronecker delta function, $N^{(\sigma)}=\sum_i \delta(\sigma_i,\sigma)$,
$(h^{(0)},\ldots,h^{(q-1)})$ is an homogeneous external field having $q$ components,
and $H_0$ is a generic $q$-states Potts Hamiltonian with no external field.
Let us rewrite $H$ as (up to terms negligible for $N\to\infty$)
\begin{eqnarray}
\label{HgPotts}
H=H_0(\{\sigma_i\})-\frac{J}{N}\sum_\sigma \left[\sum_{i}\delta(\sigma_i,\sigma)\right]^2
-\sum_{\sigma} h^{(\sigma)}N^{(\sigma)}.
\end{eqnarray} 
From Eq. (\ref{HgPotts}) we see that, by introducing $q$ independent Gaussian variables $u^{(\sigma)}$, we
can evaluate $Z|_{h^{(0)}=0,\ldots,h^{(q-1)}=0}$ as
\begin{eqnarray}
\label{Z1Potts}
&& Z\propto \int \prod_{\sigma=0}^{q-1} d u^{(\sigma)} ~e^{-NL(\{u^{(\sigma)}\})}, \quad \mathrm{where}~
L(\{u^{(\sigma)}\})= \nonumber \\
&& \sum_\sigma \frac{\beta J (u^{(\sigma)})^2}{2}+ 
\beta f_0(\beta J u^{(0)},\ldots, \beta J u^{(q-1)}),
\end{eqnarray}  
where $f_0(\beta h^{(0)},\ldots \beta h^{(q-1)})$ is the free energy density of the
Potts model governed by $H_0$ in the presence of an external field having 
$q$ components. Notice that here we are considering a generic $H_0$ which may depend on more
different couplings, therefore, from now on we will omit to write such understood dependencies. 
From Eq. (\ref{Z1Potts}) we find that, if $m_0^{(\sigma)}(\beta h^{(0)},\ldots \beta h^{(q-1)})$
is the order parameter for $H_0$ as a function of an homogeneous external field, 
$\media{\delta(\sigma_i,\sigma)}_0=m_0^{(\sigma)}(\beta h^{(0)},\ldots \beta h^{(q-1)})$, then
the order parameter for $H$, $m^{(\sigma)}=\media{\delta(\sigma_i,\sigma)}$, satisfies the system 
\begin{eqnarray}
\label{Potts}
m^{(\sigma)}=m_0^{(\sigma)}\left(\beta J m^{(0)},\ldots, \beta J m^{(q-1)}\right), \quad \forall \sigma.
\end{eqnarray}  

\begin{remark}
In conclusion, at least for the class of models so far considered, 
the general result expressed by Eqs. (\ref{m}) and (\ref{m1}) seems true.
We stress that the above derivations are rigorous and no special assumption was necessary.
However, pretending to prove Eqs. (\ref{m}) and (\ref{m1}) in all their generality
by using GT's is hopeless. In fact, if we consider \textit{e.g.} 
a Hamiltonian of the kind $H=H_0-J\sum_i\sigma_i \tanh(\sum_{j\neq i}\sigma_j/N)$ we see that
we should use an infinite number of GT's in order to prove, if any, Eqs. (\ref{m}) and (\ref{m1}).
Later we will see that, under mild conditions on the correlation functions, 
Eqs. (\ref{m}) and (\ref{m1}) can be in fact rigorously proved for the most general models.
\end{remark}

\section{Example: 1D Ising + FC 3-spins}
As an exactly solvable example, let us consider the case of Eq. (\ref{Hgp}) with
$p=3$ and with $H_0$ Ising, \textit{i.e.}, $p'=2$, in one dimension. 
For $N\to\infty$, the free energy density, 
the magnetization and the two-points CCF
for $H_0$ as a function of an homogeneous external field $h$ are given by (see \textit{e.g.}~\cite{Baxter}):
\begin{eqnarray*}
\label{1d}
&& -\beta f_0 (\beta h)= \beta J_0 + \\
&&\log\left({\cosh(\beta h)+
\left[\sinh^2(\beta h)+e^{-4\beta J_0}\right]^{\frac{1}{2}}}\right),\\
\label{1da}
&& {m_0 (\beta h)=\sinh(\beta h)
\left[\sinh^2(\beta h)+e^{-4\beta J_0}\right]^{-\frac{1}{2}}},\\
\label{1db}
&& {{C}}_{0;i,j}^{(2)} (\beta h)={\media{\sigma_i\sigma_j}}_0-{\media{\sigma}}_0^2=\sin^2(2\varphi)
e^{-|i-j|_0/\xi_0}, \nonumber \\
&& \xi_0=\log\left(\frac{\cosh(\beta h)+
\left[e^{4\beta J_0}\sinh^2(\beta h)+1\right]^{\frac{1}{2}}}
{\cosh(\beta h)-\left[e^{4\beta J_0}\sinh^2(\beta h)+1\right]^{\frac{1}{2}}}\right), \nonumber \\
&& \cot(2\varphi_0)=e^{2\beta J_0}\sinh(\beta h), \quad 0<\varphi<\frac{\pi}{2},
\end{eqnarray*} 
where ${|i-j|}_0$ is the (Euclidean) distance in 1D between $i$ and $j$.
By applying Eqs. (\ref{mpg}) and (\ref{m1}), for the system governed by $H$ with
no external field, we get, respectively
\begin{eqnarray}
\label{1daH}
&& {m ={\sinh(\beta J m^2)}
{\left[\sinh^2(\beta J m^2)+e^{-4\beta J_0}\right]^{-\frac{1}{2}}}},\\
\label{1dbH}
&& {{C}}_{i,j}^{(2)} \defi{\media{\sigma_i\sigma_j}}-{\media{\sigma}}^2=\sin^2(2\varphi)
e^{-|i-j|_0/\xi}, ~~~~~
\end{eqnarray} 
where, in Eq. (\ref{1dbH}), $\varphi=\varphi_0(\beta h\to \beta J m^2)$ and $\xi=\xi_0(\beta h\to \beta J m^2)$,
and similarly for the free energy density $\beta f=2\beta Jm^3/3 +\beta f_0(\beta h\to \beta Jm^2)$, where
$m$ is the solution of (\ref{1daH}). Notice that Eq. (\ref{1daH}) develops only first-order phase transitions
and, in this case, $p=3$, there is no negative solution.
In Fig. 1 we report the plot of $m$ as a function of $T$ obtained from Eq.~(\ref{1daH}).\\
\begin{figure}[tbh]
\epsfxsize=60mm \centerline{\epsffile{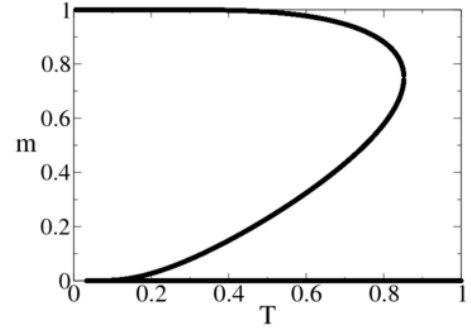}}
\caption{Solutions of Eq. (\ref{1daH}) for the case $J_0=0.2$ and $J=1$.
A bifurcation appears at $T_1\sim 0.85$, with one stable (upper curve) and one unstable
solution (lower curve). 
{The solution $m=0$ remains stable also at very low $T$ but, 
for $T<T_2$, the leading solution, \textit{i.e.}, the one having minimal free energy, 
is the upper curve while $m=0$ becomes meta-stable. Here $T_2\sim 0.77$.}}
\end{figure}

\section{General Result for Scalar Models}
\begin{definition}
Let $H_0(\{{S}_i\})$ be a Hamiltonian depending on the variables $\{{S}_i\}_{i=1}^N$ each taking values
on $\mathcal{S}$, ${S}_i\in\mathcal{S}$, with an \textit{a priori} measure $\lambda(S_i)$,
where $\mathcal{S}$ is a compact subset of 
$\mathbb{R}$. For any $\beta=1/T$, and external fields $\{{h}_i\}_{i=1}^N$, let  
\begin{eqnarray}
\label{m0}
&& {m}_{0;i}(\{{h}_i\})\defi \media{{S}_i}_0= \\
&& \frac{\int \prod_i d\lambda(S_i) {S}_i
e^{-\beta H_0(\{{S}_i\})+\sum_i \beta {h}_i\cdot {S}_i}}
{\int \prod_i d\lambda(S_i) e^{-\beta H_0(\{{S}_i\})+\sum_i \beta {h}_i\cdot {S}_i}}, \quad \mathrm{and~ similarly}
\nonumber \\
\label{C0}
&& C_{0;i,j}^{(2)}(\{{h}_i\})\defi \media{{S}_i{S}_j}_0, \quad \forall i,j=1,\ldots,N.
\end{eqnarray}
Correlation functions of higher order, $k\geq 3$, will be shortly indicated with $C_{0}^{(k)}(\{h_i\})$.
Let ${U}_i: \mathcal{S}\rightarrow \mathbb{R}$ be bounded functions of the variables $\{{S}_j\}$, and let 
\begin{eqnarray}
\label{Htheo}
H(\{{S}_i\})\defi H_0(\{{S}_i\})-\sum_i {S}_i\cdot {U}_i (\{{S}_j\}) - \sum_i {h}_i\cdot {S}_i.
\end{eqnarray}
\end{definition}

\begin{remark}
In order to have a compact notation, above and in what follows (both in the statements and in the proof
of the following theorem), we omit to write the dependence on $N$. 
For the same reason, in the formulas, we will state explicitly the $\lim N\to \infty$
only when $N$ appears explicitly.  
\end{remark}

\begin{theorem}
If, for any $N$, and for any $i$,
\begin{eqnarray}
\label{Cond1}
&&{U}_i={U}_i\left(\frac{1}{N}\sum_{j(\neq i)} {S}_j\right), \quad \mathrm{and} \\
\label{Cond2}
&&{\lim_{N\to\infty}\frac{1}{N}|\sum_{i,j}C_{i,j}^{(2)}|<\infty, \quad \forall \{{h}_i\}\neq \{{0}\}},~~~
\end{eqnarray} 
then, in the limit $N\to\infty$, for the thermal averages ${m}_i(\{{h}_i\})=\media{{S}_i}$ 
of the system (\ref{Htheo}) it holds~\footnote
{
SLLN holds even with weaker conditions than (\ref{Cond2}), see \cite{LyonsSLLN}. 
}:
\begin{eqnarray}
\label{Theomi}
{m}_{i}={m}_{0;i}(\{\beta {U}_j({m})+ \beta {h}_j\}),
\end{eqnarray}
and, similarly, for any correlation function $C^{(k)}$ of order $k$, depending on $k$ indices $i_1,\ldots,i_k$
left understood, it holds
\begin{eqnarray}
\label{TheoC}
C^{(k)}=C_0^{(k)}(\{\beta {U}_j({m})+ \beta {h}_j\}),
\end{eqnarray}
where the order parameter ${m}={m}(\{{h}_i\})$ satisfies
\begin{eqnarray}
\label{Theom}
{m}=\lim_{N\to\infty}\frac{1}{N}\sum_i{m}_{0;i}(\{\beta {U}_j({m})+ \beta {h}_j\}),
\end{eqnarray}
which, by considering an additional uniform field $h$, 
gives rise to the following susceptibility
\begin{eqnarray}
\label{TheoCchi}
\chi\defi \frac{\partial m}{\partial \beta h}=\frac{\chi_0\left(\{\beta {U}_j({m})+ \beta {h}_j\}\right)}
{1-\lim_{N\to\infty}\frac{1}{N}\sum_{p,q}C_{0;p,q}^{(2)}\frac{\partial{\beta {U}_q(m)}}{\partial m}}, 
\end{eqnarray}
where $\chi_0$ stands for the susceptibility of the system with Hamiltonian $H_0$.
For finite $N$, however,
\begin{eqnarray}
\label{TheoC1}
C^{(k)}=C_0^{(k)}(\{\beta {U}_j({m})+ \beta {h}_j\})+\mathop{O}\left(N^{-1}\right),
\end{eqnarray}
and, in particular, omitting the understood arguments,
\begin{eqnarray}
\label{TheoC2}
&& C^{(2)}=C_{0;i,j}^{(2)}+ 
\frac{1}{N}\frac{\sum_{p,q}C_{0;i,p}^{(2)}\frac{\partial{\beta {U}_p(m)}}{\partial m}C_{0;q,j}^{(2)}}
{1-\frac{1}{N}\sum_{p,q}C_{0;p,q}^{(2)}\frac{\partial{\beta {U}_q(m)}}{\partial m}}.
\end{eqnarray}

\begin{proof}
Given the Gibbs-Boltzmann distribution for the system having Hamiltonian (\ref{Htheo}),   
$p(\{{S}_i\})=\exp(-\beta H(\{{S}_i\}))/Z$,
let us derive its conditional probability $p^{(1)}({S}_i|\{{S}_j,j\neq i\})$ with respect
to the single random variable ${S}_i$.
Note that, given $H_0(\{{S}_j\})$, and an index $i$, 
it is always possible to decompose $H_0(\{{S}_j\})$ as
\begin{eqnarray}
\label{Proof}
H_0(\{{S}_j\})=H_{0;i}({S}_i|\{{S}_{j\neq i}\})+\Delta H_{0;i}(\{{S}_{j\neq i}\})
\end{eqnarray}
In other words, $H_{0;i}$ is defined as the part of $H_0(\{{S}_i\})$ that involves
the interactions of the variable ${S}_i$ with all the other variables. 
From Eqs. (\ref{Htheo}) and (\ref{Cond1}) we have
\begin{eqnarray}
\label{Proof2}
&& p^{(1)}({S}_i|\{{S}_j,j\neq i\})=\nonumber \\
&& \frac{
e^{-\beta H_{0;i}({S}_i|\{{S}_{j\neq i}\})+\beta {S}_i\cdot {U}_i ({\mathop{\rm m}})
+\beta {h}_i\cdot {S}_i}
}
{\int d(\lambda{S_i'})e^{-\beta H_{0;i}({S}_i'|\{{S}_{j\neq i}\})+\beta {S}_i'\cdot {U}_i ({\mathop{\rm m}})
+\beta {h}_i\cdot {S}_i'}},\\
\label{Proof1}
&& \mathrm{where} \quad\mathop{\rm m}\defi \frac{1}{N}\sum_{j\neq i} {S}_j.
\end{eqnarray} 
On the other hand, Eq. (\ref{Cond2}) allows to use the SLLN \cite{LyonsSLLN}~: 
\begin{eqnarray}
\label{Proof3}
\lim_{N\to \infty}\mathop{\rm m}= m, \quad \mathrm{a.s.},
\end{eqnarray} 
so that $p^{(1)}$, in the limit $N\to\infty$, can be rewritten as
\begin{eqnarray}
\label{Proof4}
&& p^{(1)}({S}_i|\{{S}_j,j\neq i\})=\nonumber \\
&& \frac{
e^{-\beta H_{0;i}({S}_i|\{{S}_{j\neq i}\})+\beta {S}_i\cdot \tilde{{h}}_i ({m})}
}
{\int d(\lambda{S_i'})e^{-\beta H_{0;i}({S}_i'|\{{S}_{j\neq i}\})+\beta {S}_i'\cdot \tilde{{h}}_i ({m})}}, \quad \mathrm{a.s.}\\
\label{Proof5}
&& \mathrm{where}\quad \tilde{{h}}_i ({m})\defi {U}_i ({m}) +{h}_i.
\end{eqnarray} 
From Eq. (\ref{Proof4}) we see that, in the limit $N\to\infty$
\begin{eqnarray}
\label{Proof6}
p^{(1)}({S}_i|\{{S}_j,j\neq i\})=\tilde{p}^{(1)}_{0}({S}_i|\{{S}_j,j\neq i\}), 
\quad \mathrm{a.s.}
\end{eqnarray} 
where $\tilde{p}^{(1)}_{0}$ is the conditional probability of the system governed by 
$H_0$ in the presence of the fields $\{\tilde{{h}}_i\}$.
Hence, from Eqs. (\ref{Proof6}) and (\ref{m0}) it follows that, in the limit $N\to\infty$
\begin{eqnarray}
\label{Proof7}
\media{{S}_i}=\media{\media{{S}_i|\{{S}_j,j\neq i\}}}
={m}_{0;i}(\{\beta {U}_j({m})+ \beta {h}_j\}).
\end{eqnarray} 
Similarly one can obtain Eqs. (\ref{TheoC}) by following very analogous steps. 
Finally, by summing Eq. (\ref{Theomi}) over the index $i$, one gets the equation for $m$, Eq. (\ref{Theom});
while Eq. (\ref{TheoC1}), and similarly Eq. (\ref{TheoC2}), are easily obtained by derivation
of Eqs. (\ref{Theomi}) and (\ref{Theom}) with respect to $\beta h_j$.
\qed
\end{proof}
\end{theorem}

For the free energy density, $f$, there is no simple expression. 
However, for the case ${U}_i\equiv {U}$ , up to an irrelevant constant, $C$, we have
\begin{eqnarray}
\label{TheoC3}
f &=& \min_{m\in \mathcal{S}} \lim_{N\to\infty}\left[N\left(m{U}(m)-\int_{\inf(\mathcal{S})}^m dm' {U}(m')\right) 
\right. \nonumber \\ && \left. 
+f_0\left(\{\beta {U}({m})+ \beta {h}_j\}\right)\right] +C.
\end{eqnarray}


\section{General Result for Vector Models}
\begin{definition}
Let $H_0(\{\bm{S}_i\})$ be a Hamiltonian depending on $n$-dimensional vectors $\{\bm{S}_i\}_{i=1}^N$ each taking values
on $\mathcal{S}$, $\bm{S}_i\in\mathcal{S}$, with an \textit{a priori} measure $\lambda(\bm{S}_i)$,
where $\mathcal{S}$ is a compact subset of 
$\mathbb{R}^n$ equipped with an internal product $\cdot:\mathcal{S}\times \mathcal{S}\rightarrow \mathbb{R}$. 
For any $\beta$, and external fields $\{\bm{h}_i\}_{i=1}^N$, let  
\begin{eqnarray}
\label{Vecm0}
&& \bm{m}_{0;i}(\{\bm{h}_i\}) \defi \media{\bm{S}_i}_0= \\
&& \frac{\int \prod_i d\lambda(\bm{S}_i) \bm{S}_i
e^{-\beta H_0(\{\bm{S}_i\})+\sum_i \beta \bm{h}_i\cdot \bm{S}_i}}
{\int \prod_i d\lambda(\bm{S}_i) e^{-\beta H_0(\{\bm{S}_i\})+\sum_i \beta \bm{h}_i\cdot \bm{S}_i}}, 
\quad \mathrm{and~ similarly} \nonumber \\
\label{VecC0}
&& C_{0;i,j}^{(2;r,s)}(\{\bm{h}_i\})\defi \media{{S}_i^{(r)}{S}_j^{(s)}}_0, \quad \forall i,j=1,\ldots,N.
\end{eqnarray}
Correlation functions of higher order, $k\geq 3$, will be also shortly indicated by $\bm{C}_{0}^{(k)}(\{\bm{h}_i\})$.
Note that, for any choice of $k$ different indices $i_1,\ldots,i_k$, 
$\bm{C}_{0;i_1,\ldots,i_k}^{(k)}(\{\bm{h}_i\})$ represents a $(n\times n \times \cdots \times n)$ tensor of rank $k$.
Let $\bm{U}_i: \mathcal{S}\rightarrow \mathbb{R}^n$ bounded functions of the vectors $\{\bm{S}_j\}$, and let 
\begin{eqnarray}
\label{VecHtheo}
H(\{\bm{S}_i\})= H_0(\{\bm{S}_i\})-\sum_i \bm{S}_i\cdot \bm{U}_i (\{\bm{S}_j\}) - \sum_i \bm{h}_i\cdot \bm{S}_i.
\end{eqnarray}
\end{definition}

\begin{theorem}
If, for any $i$,
\begin{eqnarray}
\label{VecCond1}
&&\bm{U}_i=\bm{U}_i\left(\frac{1}{N}\sum_{j(\neq i)} \bm{S}_j\right), \quad \mathrm{and} \\
\label{VecCond2}
&&{\lim_{N\to\infty}\frac{1}{N}|\sum_{i,j}\sum_{r,s}C_{i,j}^{(2;r,s)}|<\infty, 
\quad \forall \{\bm{h}_i\}\neq \{\bm{0}\}},~~~
\end{eqnarray} 
then, in the limit $N\to\infty$, for the thermal averages $\bm{m}_i(\{\bm{h}_i\})=\media{\bm{S}_i}$ 
of the system (\ref{Htheo}) it holds:
\begin{eqnarray}
\label{VecTheomi}
\bm{m}_{i}=\bm{m}_{0;i}(\{\beta \bm{U}_j(\bm{m})+ \beta \bm{h}_j\}),
\end{eqnarray}
and, similarly, for any correlation function $\bm{C}^{(k)}$ of order $k$, depending on $k$ indices $i_1,\ldots,i_k$,
left understood
\begin{eqnarray}
\label{VecTheoC}
\bm{C}^{(k)}=\bm{C}_0^{(k)}(\{\beta \bm{U}_j(\bm{m})+ \beta \bm{h}_j\}),
\end{eqnarray}
where the order parameter $\bm{m}=\bm{m}(\{\bm{h}_i\})$ satisfies
\begin{eqnarray}
\label{VecTheom}
\bm{m}=\lim_{N\to\infty}\frac{1}{N}\sum_i\bm{m}_{0;i}(\{\beta \bm{U}_j(\bm{m})+ \beta \bm{h}_j\}),
\end{eqnarray}
which gives rise to the following susceptibility
\begin{eqnarray}
\label{VecTheoChi}
\bm{\chi}= 
\left(1-\lim_{N\to\infty}\frac{1}{N}\sum_{p,q}\bm{C}_{0;p,q}^{(2)}\cdot\frac{\partial{\beta \bm{U}_q(\bm{m})}}{\partial \bm{m}}\right)^{-1}
\cdot\bm{\chi}_{0},
\end{eqnarray}
where $\bm{\chi}^{(r,s)}=((\partial \bm{m}/\partial \bm{\beta h}))^{(r,s)}=
\partial m^{(r)}/\partial \beta h^{(s)}$,
$(({\partial{\beta \bm{U}_{p}(\bm{m})}}/{\partial \bm{m}}))^{(r,s)}=
{\partial{\beta {U}_{p}^{(r)}(\bm{m})}}/{\partial {m}^{(s)}}$, and the products are understood as matrix products.
For finite $N$
\begin{eqnarray}
\label{VecTheoC1}
\bm{C}^{(k)}=\bm{C}_0^{(k)}(\{\beta \bm{U}_j(\bm{m})+ \beta \bm{h}_j\})+\mathop{O}\left(N^{-1}\right),
\end{eqnarray}
and, in particular, omitting for shortness the understood arguments, for the tensor $\bm{C^{(2)}}$ (a matrix), it holds
\begin{eqnarray}
\label{VecTheoC2}
&& \bm{C}_{0;i,j}^{(2)}=\bm{C}_{0;i,j}^{(2)}+
\left(1-\frac{1}{N}\sum_{p,q}\bm{C}_{0;p,q}^{(2)}\cdot\frac{\partial{\beta \bm{U}_q(\bm{m})}}{\partial \bm{m}}\right)^{-1}
\nonumber \\ && \times 
\left(
\frac{1}{N}\sum_{p,q}\bm{C}_{0;i,p}^{(2)}\cdot\frac{\partial{\beta \bm{U}_{p}(\bm{m})}}
{\partial \bm{m}}\cdot\bm{C}_{0;q,j}^{(2)}
+\frac{1}{N^2} \right.
\nonumber \\ && \times
\left.
\left[
\sum_{p'}\bm{C}_{0;i,p'}^{(2)}\cdot\frac{\partial{\beta \bm{U}_{p'}(\bm{m})}}{\partial \bm{m}},
\sum_{l,p}\bm{C}_{0;l,p}^{(2)}\cdot\frac{\partial{\beta \bm{U}_{p}(\bm{m})}}{\partial \bm{m}}
\right]
\right.
\nonumber \\ && \times
\left.
\sum_{q}\bm{C}_{0;q,j}^{(2)} 
\right),
\end{eqnarray}
where $\left[\bm{A},\bm{B}\right]=\bm{A}\bm{B}-\bm{B}\bm{A}$. 
Note that the commutator in (\ref{VecTheoC2}) is zero only in the case of an homogeneous system,
however, the sum over the index $i$ of such a commutator gives always zero. Similarly, by using another
analogous expression, by summing over $j$ the corresponding commutator gives zero.
This fact ensures that by summing $\bm{C}_{0;i,j}^{(2)}$ over $i$ and $j$ gives back, consistently, Eq. (\ref{VecTheoChi}).

\begin{proof}
The proof runs almost exactly as in the scalar case with the obvious dependencies on the 
vector indices. With respect to the scalar case, the only main difference lies 
in Eq. (\ref{VecTheoC2}) for which the presence of a non zero commutator can make, for finite $N$,
the correlation of two components of two given vectors quite larger then the correlation
of two given scalars, as given by Eq. (\ref{TheoC2}).
\qed
\end{proof}
\end{theorem}

\section{Conclusions}
Theorems 1 or 2 prove the generalized mean-field Eqs. (\ref{m}) and (\ref{m1})
(including the presence of a non uniform external field),
for scalar or vector models, respectively, and provide the exact expression for the 
CCF (including finite size terms of order $\mathop{O}(1/N)$) that, unlike ordinary mean-field models,
persist even in the thermodynamic limit, and that are in turn responsible for a rich scenario of phase transitions;
\textit{e.g.}, if $H_0$ has negative couplings, first-order and inverse transition phenomena 
can occur even in the Ising case.
Theorems 1 or 2 are valid upon the mild bounds (\ref{Cond2}) or (\ref{VecCond2}) 
on the two-points CCF for $H$. Typically, if the two-points CCF decay exponentially fast with
the distance, the conditions (\ref{Cond2}) or (\ref{VecCond2}) hold.
Note that, in a pure Gaussian theory, all CCF 
of order higher than 2 are zero but, in our case, 
we may have non zero CCF at any order. 
Whereas, in general, is mathematically hard to prove,
\textit{a priori}, 
the conditions (\ref{Cond2}) or (\ref{VecCond2}),
such bounds, in physical terms, just mean that the system $H$ - and then necessarily
the system $H_0$ - must have a finite susceptibility up to a domain of zero measure
in the space of the parameters: the critical surface. It is a matter of fact that most of the systems
with finite range interactions satisfy this requirement. A sort of exception is the Ising model
on the Cayley tree (not to be confused with the Bethe lattice) where, at zero external field, 
the susceptibility diverges on an entire interval of temperatures \cite{Muller}.
But, as is known, this effect is due to the non amenable structure of the Cayley tree which
heavily depends on the boundary conditions. However,
even in this extreme case, the susceptibility remains finite for any non zero field.
A similar situation occurs in 2D for several models having continuous symmetries, like the XY model:
the CCF are power-law, but only for zero field.  
Perhaps, the only actual counter-examples are the models built on networks with a fat tail degree distribution.
In this case, the CCF remain strong (\textit{i.e.} power law) 
even for non zero field (and also for any $T$) \cite{SWSF}. For such models then, we cannot
apply Theorems 1 or 2. However, as we have shown in the Second Section, if $\Delta H$ is sufficiently ``simple'', 
we can still apply the GT's, as it has been in fact done in \cite{SWSF} within the context of 
scale-free small-world networks. 

{We thank J. H. H. Perk for pointing out to us Refs. \cite{Bogoliubov}
and \cite{Perk} and for useful discussions.
Work supported by Grant IIUM EDW B 11-159-0637}.


\newpage
\section{Supplementary Material}
The methods M1 and M2 mentioned in the Introduction,
can be similarly used to solve exactly the $p$-spins model built on the FC graph.  
In this case each interaction is the product of $p\geq 2$ spins:
\begin{eqnarray}
\label{Hpspin}
H_p\defi -\frac{(p-1)!J}{N^{p-1}}\sum_{i_1<\ldots<i_p} \sigma_{i_1}\cdots \sigma_{i_p}+h\sum_i\sigma_i,
\end{eqnarray}
and, in the thermodynamic limit, $m$ satisfies
\begin{eqnarray}
\label{mpspin}
m=\tanh(\beta J m^{p-1} +\beta h).
\end{eqnarray}
From Eq. (\ref{mpspin}) it follows that, when $p>2$, the transitions are of first-order and, furthermore, 
at each temperature Eq. (\ref{mpspin}) may have more independent stable solutions, and
the thermodynamic leading solution is the one which minimizes, for $m\in(-1:1)$, the following Landau functional:
\begin{eqnarray}
\label{tLpspin}
{L}_2(m)=-\frac{\beta J m^p}{p}+\beta h m+g(m),
\end{eqnarray}
where $g$ is defined as $g(m)=-\log[\mathcal{N}(Nm)]/N$ and, up to terms $\mathop{O}(1/N)$, is given by
$g(m)=(1-m)/2\log\left((1-m)/2\right)+(1+m)/2\log\left((1+m)/2\right)$.
More precisely, Eq. (\ref{tLpspin}) comes from the use of M2, while the use of M1
gives the following Landau functional:
\begin{eqnarray}
\label{Lpspin}
L_1(m)=\frac{\beta J (p-1)m^p}{p}-\log\left[2\cos(\beta Jm^{p-1}+\beta h)\right].
\end{eqnarray}
Note that $L_1(m)$ and ${L}_2(m)$ coincide, and provide the free energy density $f$ 
of the system, only when calculated in correspondence of their (common) local or global minima:
\begin{eqnarray}
\label{fpspin}
\beta f=\min_{m\in(-1:1)} L_1(m)=\min_{m\in(-1:1)} {L}_2(m).
\end{eqnarray}

\end{document}